\documentclass[aps,twocolumn,prd,superscriptaddress]{revtex4-2}
\usepackage{amssymb}
\usepackage{amsmath} 
\usepackage{slashed}
\usepackage{orcidlink}
\usepackage{graphicx}
\usepackage{extpfeil}
\usepackage{babel}
\usepackage{hyperref}
\usepackage{color}
\usepackage{comment,bm}

\newcommand{\dd}{\mathrm{d}}
\newcommand{\cN}{\mathcal{N}}
\newcommand{\cL}{\mathcal{L}}
\newcommand{\cE}{\mathcal{E}}
\newcommand{\cU}{\mathcal{U}}
\newcommand{\cQ}{\mathcal{Q}}
\newcommand{\cF}{\mathcal{F}}
\newcommand{\Del}{\Delta}

\begin{document}

\title{Half-BPS Impurity Backgrounds and Supersymmetry}

\author{D. Bazeia\,\orcidlink{0000-0003-1335-3705}}\email[]{bazeia@fisica.ufpb.br}\affiliation{Departamento de F\'\i sica, Universidade Federal da Para\'\i ba, 58051-970 Jo\~ao Pessoa, PB, Brazil}

\author{A. C. Lehum\,\orcidlink{0000-0001-5075-6541}}\email[]{lehum@ufpa.br}
\affiliation{Faculdade de F\'{i}sica, Universidade Federal do Par\'{a}, 66075-110, Bel\'{e}m, Par\'a, Brazil}


\begin{abstract}
We develop a rigid $\mathcal{N}=(1,1)$ superspace framework for spatially inhomogeneous impurity deformations in $D=1+1$ dimensions by embedding the impurity profile into a real background superfield (spurion). This spurionic completion provides a manifestly supersymmetric description at the level of the action and offers a systematic route to identify which inhomogeneous backgrounds preserve a nontrivial subset of supercharges. Focusing on static interface-type configurations, we determine the half-BPS condition on the spurion background and the corresponding supersymmetry projector. In the resulting half-BPS sector we derive the associated first-order BPS equation for static bosonic matter configurations and establish an exact Bogomol'nyi completion of the static energy, yielding a sharp bound saturated by BPS solutions. We further comment on how explicit coordinate dependence and derivative-dependent impurity couplings can obstruct the Bogomol'nyi structure, thereby motivating spurionic extensions that retain supersymmetric control over inhomogeneous deformations.

\end{abstract}

\maketitle

{{\bf I.} {\bf \textit {Introduction.}}} Impurity deformations, understood as prescribed spatial inhomogeneities that couple to the order parameter, have long been known to modify the spectrum and interactions of solitons and defects \cite{Malomed,impu}. A recent and interesting possibility has appeared in Ref. \cite{Adam}, in which the spectral wall phenomenon was identified. In the absence of impurities, the study of localized structures may be directly related to the work of Bogomol'nyi \cite{B} and Prasad and Sommerfield \cite{PS}, referred to as the BPS framework. However, the presence of impurities changes the picture, and in recent years, a systematic ``BPS-preserving'' impurity program has been developed in non-supersymmetric settings, where suitable impurity couplings lead to first-order equations and a Bogomol'nyi-type bound for selected topological sectors, providing a controlled framework to study localized structures and impurity interactions in high energy physics; see, e.g., Refs.~\cite{Bazeia:2023rug,Bazeia:2023dyr} and references therein. 

A direct supersymmetric counterpart of the BPS impurity idea in $1+1$ spacetime dimensions was formulated by Adam, Queiruga, and Wereszczy\'nski \cite{Adam:2019yst}, who exhibited soliton-impurity models that preserve \emph{half} of the rigid supersymmetry and in which only one orientation of the topological defect remains BPS. Their construction makes precise the notion that a spatially dependent impurity generically breaks supersymmetry, but a nontrivial subset of supercharges can survive when the impurity is embedded into an enlarged structure that admits a consistent Bogomol'nyi completion and a projected supersymmetry algebra.

Motivated by the results of Ref. \cite{Adam:2019yst} and by the BPS-impurity formalism explored in the non-supersymmetric literature \cite{Bazeia:2023rug,Bazeia:2023dyr}, we revisit the problem within a framework designed to interface directly with the impurity couplings introduced in Ref.~\cite{Bazeia:2023dyr}, while keeping the analysis fully transparent at the level of superspace; See Refs. \cite{GGRS,Shizuya:2003vm} and the Appendix  for more on supersymmetry. Concretely, we promote the impurity profile to a real spurion superfield $\Sigma$ and work with a general class of couplings of the schematic form $\Sigma\,\cF(\Phi)$ supplemented by the minimal quadratic completion in $\Sigma$. 
This strategy allows us to derive the half-BPS condition on the impurity background as a purely algebraic requirement from $\delta\lambda_\pm=0$, identify explicitly the surviving supercharge via a projector on supersymmetry parameters, and obtain the corresponding universal first-order equation for the matter scalar, together with an \emph{exact} Bogomol'nyi decomposition of the static energy into a perfect square plus a total derivative. The resulting framework provides a clean bridge between the supersymmetric half-BPS impurity mechanism \cite{Adam:2019yst} and the broad class of impurity-doped models studied in the non-supersymmetric program \cite{Bazeia:2023rug,Bazeia:2023dyr}, and it is well suited for systematic generalizations, as in the case of multi-field sectors, derivative-dependent impurity functionals, and gauge/vortex/monopole extensions in higher dimensions.

We start the investigation in Sec.~II, presenting the spurion-completed superspace action and deriving the full component Lagrangian together with the component supersymmetry transformations for both the matter and spurion multiplets. In Sec.~III we impose a static bosonic spurion background and determine the necessary and sufficient condition for preserving a projected supercharge, as well as the associated half-BPS projector; we then derive the universal first-order BPS equation for static bosonic matter configurations. In Sec.~IV we demonstrate that, once the half-BPS spurion condition holds, the static energy admits an exact Bogomol'nyi completion into a perfect square plus a total derivative, leading to a sharp bound saturated precisely by BPS solutions. In Secs.~V and VI we discuss extensions in which the impurity functional carries explicit $x$-dependence or depends on superspace/spacetime derivatives, clarifying when and how the Bogomol'nyi structure is obstructed and how it can be restored by enlarging the spurion sector. We summarize our results and outline future directions in Sec.~VII.

{{\bf II.} \bf{\textit{Superspace action.}}}
\label{sec:action}
We consider the spurion-completed superspace action
\begin{eqnarray}
S=\int \dd^2x\,\dd^2\theta\Bigl(\!&-&\!\frac12\,(D_\alpha\Phi)^2- W(\Phi) \nonumber\\
&-& \Sigma\,\cF(\Phi,\Sigma)
-\frac12\,\Sigma^2
\Bigr),
\label{eq:action-superspace}
\end{eqnarray}
with arbitrary real superfunctions $W(\Phi)$ and $\cF(\Phi,\Sigma)$; See the Apendix for superspace notation and conventions. The real scalar superfield $\Phi$ (matter multiplet) and the real spurion superfield $\Sigma$ (background multiplet) are expanded as
\begin{align}
\Phi(x,\theta)\! &=\! \phi(x)\!-\theta_+\psi^+(x)-\theta_-\psi^-(x)
-i\theta^+\theta^- F(x),
\label{eq:Phi-exp}
\\
\Sigma(x,\theta)\! &=\! \sigma(x)\!
-\theta_+\lambda^+(x)
-\theta_-\lambda^-(x)
-i\theta^+\theta^-\,G(x),
\label{eq:Sigma-exp}
\end{align}
where $\psi^\pm$ and $\lambda^\pm$ are real Majorana--Weyl fermions and $F,G$ are real auxiliary scalars.

When $\Sigma$ is treated as a fixed background superfield, the action remains manifestly supersymmetric at the level of superspace. Supersymmetry is \emph{explicitly} broken only upon freezing $\Sigma$ to a prescribed purely bosonic profile $\sigma(x)$ and setting its superpartners to zero. In this sense, our construction provides a $D=1+1$ superfield extension in flat spacetime of the impurity model developed in Ref. \cite{Bazeia:2023dyr}.

\subsection*{II.A. {Component expansion}}
\label{sec:components}

\subsubsection*{Kinetic term}

Using \eqref{eq:Q-D} and \eqref{eq:Phi-exp}, one finds
\begin{align}
D_\alpha\Phi &= \psi_\alpha + \theta^\beta (i\partial_{\alpha\beta}\phi+C_{\alpha\beta} F)+\theta^2\, i{\partial_{\alpha}}^{\beta}\psi_\beta,
\label{eq:DplusPhi}
\\
D^\alpha\Phi &= \psi^\alpha + \theta^\beta (i{\partial_{\beta}}^{\alpha}\phi-\delta^\alpha_{\beta} F)
-\theta^2\, i{\partial^{\alpha}}_{\beta}\psi^\beta.
\label{eq:DminusPhi}
\end{align}
Multiplying and extracting the $\theta^2$ coefficient yields
\begin{eqnarray}
\big[D^\alpha\Phi\,D_\alpha\Phi\big]_{\theta^2}
=2\left[ \phi\Box\phi+\psi^\alpha i{\partial_\alpha}^\beta \psi_\beta+F^2\right]
.
\label{eq:kin-theta-comp}
\end{eqnarray}
Using \eqref{eq:berezin}, we see that the superspace contribution $-\frac12\int \dd^2\theta\, (D_\alpha\Phi)^2$ gives
\begin{equation}
\cL_{\rm kin}
=
\frac12 \left[\phi\Box\phi
-i\,\psi^+ {\partial_{++}}\psi^+
-i\,\psi^- {\partial_{--}}\psi^-
+F^2\right].
\label{eq:Lkin}
\end{equation}

\subsubsection*{Superpotential and impurity terms}

Expanding $W(\Phi)$ and extracting the $\theta^2$ component yields
\begin{equation}
\big[W(\Phi)\big]_{\theta^2}
=F\,W_\phi(\phi)+i\, W_{\phi\phi}(\phi)\,\psi^+\psi^-,
\label{eq:Wtheta}
\end{equation}
where $W_\phi(\phi)\equiv {dW(\phi)}/{d\phi}$, $W_{\phi\phi}(\phi)\equiv {d^2W(\phi)}/{d\phi^2}$ and so on. Similarly,
\begin{eqnarray}
\big[\Sigma\,\cF(\Phi,\Sigma)\big]_{\theta^2}\!\!
&=&
G\cF+\sigma\Big[\cF_\phi\,F +\cF_\sigma G
\nonumber\\
&&+ i\cF_{\phi\phi}\,\psi^+\psi^-
+i\cF_{\sigma\sigma}\,\lambda^+\lambda^-
\nonumber\\
&&
+i\cF_{\phi\sigma}(\psi^+ \lambda^- - \psi^- \lambda^+)\Big]\nonumber\\
&&+i\,\cF_\phi (\lambda^+ \psi^- - \lambda^-\psi^+)
\nonumber\\
&&+2i\cF_\sigma \lambda^+ \lambda^-
,
\label{eq:SigKtheta}
\end{eqnarray}
with $\cF=\cF(\phi,\sigma)$ at component level, and
\begin{equation}
\Big[\frac12\,\Sigma^2\Big]_{\theta^2}
=
2\sigma\,G+i\lambda^+\,\lambda^-.
\label{eq:Sig2theta}
\end{equation}

\subsubsection*{Full off-shell Lagrangian}

Collecting Eqs.~\eqref{eq:Lkin}--\eqref{eq:Sig2theta} and substituting them into the superspace action \eqref{eq:action-superspace}, we obtain the off-shell component Lagrangian
\begin{eqnarray}
\cL_{\rm off}
&=&\;
\frac12\,\phi\,\Box\phi
+\frac12\,F^2
-\;F\Big(W_{\phi}+\sigma\,\cF_{\phi}\Big)
\nonumber\\
&& -\;G\Big(\cF+\sigma+\sigma\cF_\sigma\Big)+\ldots,
\label{eq:Loff-full}
\end{eqnarray}
where the ellipsis denotes the terms involving the fermionic fields. 

At this point, it is worth stressing an important difference with respect to Ref.~\cite{Bazeia:2023dyr}. In the present superspace formulation, the quadratic spurion term $\Sigma^2$ produces, in components, a term proportional to $G\,\sigma$, rather than a pure $\sigma^2$ contribution. This is a direct consequence of embedding the impurity into a full background superfield. As will be shown below, after imposing the half-BPS condition $G=\eta\,\sigma'(x)$, this term becomes a total derivative and hence contributes only through a surface term that vanishes for localized impurity profiles. However, this contribution may become physically relevant in more general defect-like backgrounds, particularly when the impurity is replaced by an extended interface or by a profile with nontrivial asymptotic behavior.

Varying with respect to $F$ gives the algebraic equation of motion
\begin{equation}
F=W_\phi(\phi)+\sigma\,\cF_\phi(\phi,\sigma).
\label{eq:F-eom}
\end{equation}

\subsection*{II.B. Component supersymmetry variations}
\label{sec:variations}

Using \eqref{eq:Q-D} and \eqref{eq:Phi-exp}--\eqref{eq:Sigma-exp}, the component variations are:

\subsubsection*{Matter multiplet}

\begin{align}
\delta\phi &= -i(\varepsilon^+\psi^- - \varepsilon^-\psi^+),
\label{eq:delta-phi}
\\
\delta\psi_+ &= i(\varepsilon^- F -\varepsilon^+\,\partial_{++}\phi),
\label{eq:delta-psi+}
\\
\delta\psi_- &= -i(\varepsilon^+ F +\,\varepsilon^-\,\partial_{--}\phi),
\label{eq:delta-psi-}
\\
\delta F &= i(\varepsilon^+\,\partial_{++}\psi^+ + \varepsilon^-\,\partial_{--}\psi^-),
\label{eq:delta-F}
\end{align}

\subsubsection*{Spurion multiplet}

\begin{align}
\delta\sigma &= -i(\varepsilon^+\lambda^- - \varepsilon^-\lambda^+),
\label{eq:delta-sigma}
\\
\delta\lambda_+ &= i(\varepsilon^- G -\varepsilon^+\,\partial_{++}\sigma),
\label{eq:delta-lp}
\\
\delta\lambda_- &= -i(\varepsilon^+ G +\varepsilon^-\,\partial_{--}\sigma),
\label{eq:delta-lm}
\\
\delta G &= i(\varepsilon^+\,\partial_{++}\lambda^+ + \varepsilon^-\,\partial_{--}\lambda^-),
\label{eq:delta-G}
\end{align}

At this stage we have obtained the complete off-shell component formulation of the matter and spurion multiplets, together with their rigid $\mathcal{N}=(1,1)$ supersymmetry transformations. We can now turn to the central question addressed in the next section: identifying the static spurion backgrounds that preserve a projected supercharge and deriving the corresponding BPS reduction for the matter sector. To this end, we will restrict to bosonic, time-independent impurity profiles and require the existence of constant supersymmetry parameters for which the fermionic variations vanish. This condition fixes the half-BPS spurion background and simultaneously induces a supersymmetry projector. Implementing the same projector in the matter variations then leads to a first-order BPS equation for static bosonic configurations and sets the stage for the Bogomol'nyi analysis of the energy.

{{\bf III.} {\bf \textit{Half-BPS spurion background and matter BPS equation.}}}
\label{sec:BPS}
It is instructive to contrast our spurion-based construction with the half-supersymmetric impurity framework of Ref.~\cite{Adam:2019yst}. There, the impurity deformation is formulated at the component level so that the Lagrangian is invariant only under a restricted subset of supersymmetry transformations. In that setting, the preserved supercharge is most conveniently exhibited by working in a basis for the Majorana parameter in which one component is set to zero, with the other left free. In our formulation, by contrast, we start from a manifestly $\mathcal{N}=(1,1)$ supersymmetric superspace action in which the impurity is encoded in a real background superfield (a spurion). Half-supersymmetry then emerges once the spurion is frozen to a static bosonic profile and one requires the existence of nontrivial supersymmetry parameters for which the fermionic spurion variation vanishes. This requirement produces a projector relating the two components of the supersymmetry parameter, thereby selecting a one-dimensional subspace of preserved transformations. Upon a change of basis in parameter space, this projector becomes equivalent to the familiar condition that one component vanishes, showing that the two descriptions capture the same notion of a single preserved supercharge while emphasizing complementary viewpoints. With this perspective in mind, we now determine the half-BPS spurion background and derive the corresponding BPS equation for static bosonic matter configurations, showing that the two descriptions identify the same notion of ``one preserved supercharge'' while emphasizing complementary perspectives: an \emph{a priori} reduced invariance at the component level versus a manifestly supersymmetric superspace embedding with half-BPS preservation arising from the chosen background.

\subsection*{III.A. Half-BPS spurion condition}

We consider a static bosonic spurion background
\begin{equation}
\lambda_\pm=0,
\qquad
\sigma=\sigma(x),
\qquad
G=G(x).
\label{eq:spurion-bkgd}
\end{equation}
Staticity implies, via \eqref{eq:static-ppmm},
\begin{equation}
\partial_{++}\sigma=\sigma'(x),
\qquad
\partial_{--}\sigma=-\sigma'(x),
\label{eq:static-sigma}
\end{equation}
where $\sigma'(x)=d\sigma/dx$. Preserving supersymmetry on a bosonic background requires $\delta\lambda_\pm=0$ for some nonzero constant parameters
$(\varepsilon^+,\varepsilon^-)$.
Using \eqref{eq:delta-lp}--\eqref{eq:delta-lm} and \eqref{eq:static-sigma} we obtain
\begin{align}
0 &= -i\,[\varepsilon^+\,\sigma'(x) - \varepsilon^-\,G(x)],
\label{eq:spurion-eq1}
\\
0 &=-i[\varepsilon^+\,G(x) - \varepsilon^-\,\sigma'(x)].
\label{eq:spurion-eq2}
\end{align}
These are homogeneous linear relations for $(\varepsilon^+,\varepsilon^-)$. A nontrivial solution exists if the determinant vanishes, which enforces
\begin{equation}
G(x)^2=\sigma'(x)^2
\;\Longrightarrow\;
G(x)=\pm\sigma'(x).
\label{eq:Gpm}
\end{equation}

Equation~\eqref{eq:Gpm} shows that, in the presence of a spatial gradient $\sigma'(x)\neq0$, the auxiliary component $G(x)$ of the spurion cannot be set to zero if one wishes to preserve any supersymmetry. Rather, $G$ must be switched on so as to compensate the obstruction generated by $\partial_x\sigma$ in the fermionic variations $\delta\lambda_\pm$, allowing for a nontrivial solution for the projected supersymmetry parameters. This is the superspace counterpart of the ``compensating'' impurity couplings to auxiliary fields introduced in Ref.~\cite{Adam:2019yst}, where terms linear in the auxiliary sector are required to cancel the $\sigma'(x)$-dependent contributions (up to total derivatives) and thereby retain a half-BPS subsector.

We parameterize the two branches by a sign $\eta=\pm1$:
\begin{equation}
G(x)=\eta\,\sigma'(x),
\qquad
\eta=\pm1.
\label{eq:G-eta}
\end{equation}
Substituting \eqref{eq:G-eta} into \eqref{eq:spurion-eq1} gives the projector on supersymmetry parameters,
\begin{equation}
\varepsilon^+\,\sigma' - \varepsilon^-\,\eta\,\sigma'=0
\;\Longrightarrow\;
\;\varepsilon^- = \eta\,\varepsilon^+.
\label{eq:eps-projector}
\end{equation}
Equivalently, the preserved supercharge is the linear combination
\begin{equation}
\cQ_\eta \equiv Q_+ + \,\eta\,Q_-,
\qquad
\delta_\eta=\varepsilon^+\,\cQ_\eta.
\label{eq:preservedQ}
\end{equation}
Thus, a spatially varying impurity can preserve \emph{half} of $\cN=(1,1)$ supersymmetry provided its auxiliary partner satisfies \eqref{eq:G-eta}; the surviving symmetry is generated by \eqref{eq:preservedQ}.

\subsection*{III. B. Matter BPS equation}

For a bosonic static matter configuration
\begin{equation}
\psi_\pm=0,
\qquad
\phi=\phi(x),
\qquad
F=F(x),
\label{eq:matter-bkgd}
\end{equation}
preservation of the same supercharge requires $\delta\psi_\pm=0$ with the projector \eqref{eq:eps-projector}. Using \eqref{eq:delta-psi+} with $\partial_{++}\phi=\phi'$ gives
\begin{eqnarray}
&&0=\delta\psi_+=i(\varepsilon^- F -\varepsilon^+\,\phi')
= i\,\varepsilon^+(\eta\,F - \,\phi')\nonumber\\
&&\Longrightarrow\,
F=\eta\,\phi'.
\label{eq:F-phi}
\end{eqnarray}
Combining \eqref{eq:F-phi} with the auxiliary equation of motion \eqref{eq:F-eom} yields the universal first-order (BPS) equation
\begin{equation}
\phi'(x)=\eta\Big(W_{\phi}(\phi)+\sigma(x)\,\cF_{\phi}(\phi,\sigma)\Big),
\qquad \eta=\pm1.
\label{eq:BPS-eq}
\end{equation}

We are now in a position to turn from the supersymmetry analysis to the energetics of the static sector. In the next section we show how the half-BPS spurion condition and the corresponding projector lead to a Bogomol'nyi completion of the energy and to the associated bound, saturated by solutions of the BPS equation.

{{\bf IV.} {\bf \textit{Exact Bogomol'nyi completion and bound.}}}
\label{sec:bound}
We now show that, upon imposing the half-BPS spurion condition \eqref{eq:G-eta}, the static energy admits an exact Bogomol'nyi decomposition.

\subsection*{IV.A. Energy density with auxiliaries eliminated}

Let us restrict to the bosonic sector $\psi_\pm=\lambda_\pm=0$ and static fields.
From \eqref{eq:Loff-full} and \eqref{eq:static-ppmm}, the static Lagrangian density reads
\begin{eqnarray}
\cL_{\rm off}\Big|_{\rm static}
&=&
-\frac12\,\phi'^2
+\frac12\,F^2
-\;F\Big(W_\phi+\sigma \cF_\phi\Big)
\nonumber\\
&& -\;G\Big(\cF+\sigma+\sigma\cF_\sigma\Big),
\label{eq:Lstatic}
\end{eqnarray}
so the static energy density is $\cE=-\cL_{\rm off}|_{\rm static}$,
\begin{eqnarray}
\cE
&=&
\frac{1}{2}\,\phi'^2
-\frac{1}{2}\,F^2
+\;F\Big(W_\phi+\sigma \cF_\phi\Big)
\nonumber\\
&& +\;G\Big(\cF+\sigma+\sigma\cF_\sigma\Big).
\label{eq:E-pre}
\end{eqnarray}
Eliminating $F$ via \eqref{eq:F-eom} gives
\begin{equation}
\cE
=
\frac{1}{2}\,\phi'^2
+\frac12\Big(W_\phi+\sigma \cF_\phi\Big)^2
+\;G(x)\Big(\cF+\sigma+\sigma\cF_\sigma\Big).
\label{eq:E-Felim}
\end{equation}
Imposing the half-BPS spurion background \eqref{eq:G-eta}, i.e.\ $G=\eta\sigma'$, yields
\begin{eqnarray}
\cE
&=&
\frac12\,\phi'^2
+\frac12\Big(W_\phi+\sigma \cF_\phi\Big)^2\nonumber\\
&&+\eta\,\sigma'(x)\Big(\cF+\sigma+\sigma\cF_\sigma\Big).
\label{eq:E-BPSspurion}
\end{eqnarray}

To carry out the Bogomol'nyi completion, let us first recall Eq.~\eqref{eq:F-eom},
$F= W_\phi(\phi)+\sigma(x)\,\cF_\phi(\phi,\sigma)$. 
It is also convenient to introduce the boundary functional
\begin{equation}
\cU(\phi,\sigma)\equiv W(\phi)+\sigma\,\cF(\phi,\sigma)+\frac12\,\sigma^2,
\label{eq:U-def}
\end{equation}
for which one finds the exact relation
\begin{equation}
\frac{\dd\cU}{\dd x}
=
\phi'\,F+\sigma'(x)\big(\cF+\sigma+\sigma\cF_\sigma\big).
\label{eq:total-derivative}
\end{equation}
Using
\begin{equation}
\frac12\Big(\phi'-\eta F\Big)^2
=
\frac12\,\phi'^2+\frac12\,F^2-\eta\,\phi'\,F,
\label{eq:square}
\end{equation}
and combining \eqref{eq:E-BPSspurion} with \eqref{eq:total-derivative} gives the exact Bogomol'nyi form
\begin{eqnarray}
\cE
&=&
\frac{1}{2}\Big(\phi'-\eta\big[W_\phi(\phi)+\sigma \cF_\phi(\phi,\sigma)\big]\Big)^2\nonumber \\
&+&\eta\,\frac{\dd}{\dd x}\Big(W(\phi)+\sigma \cF(\phi,\sigma)+\tfrac12\sigma^2\Big).
\label{eq:bogomolnyi}
\end{eqnarray}
The bound follows by integrating over $x$ and using the manifest positivity of the square:
\begin{equation}
E\equiv\int_{-\infty}^{+\infty}\dd x\,\cE
\ \ge\
\left|\Del \cU \right|,
\label{eq:bound}
\end{equation}
where $\Del\, \cU \equiv \cU(+\infty)-\cU(-\infty)$.
The bound is saturated if the square in \eqref{eq:bogomolnyi} vanishes, which is precisely the BPS equation \eqref{eq:BPS-eq}. The sign choice $\eta=\pm1$ is fixed by the topological sector (i.e.\ by the sign of $\Del\,\cU$) so that the boundary contribution appears as $+|\Del\,\cU|$.

{{\bf V.} {\bf \textit{Explicit spatial dependence in the impurity functional.}}}
\label{sec:K-explicit-x}
In the construction above the impurity coupling is encoded by a local function $\cF(\Phi,\Sigma)$, so that all inhomogeneity enters through the background profile of the spurion superfield $\Sigma$. One may ask whether the same half-BPS mechanism survives if the impurity functional depends \emph{explicitly} on the spatial coordinate, i.e.\ if one replaces
$ 
\cF(\Phi,\Sigma)\longrightarrow \cF(\Phi,\Sigma;x).
\label{eq:K-explicit}
$
This modification is qualitatively different. The reason is that rigid supersymmetry acts not only on the superfields but also, through the translation generators contained in $Q_\pm$, on any explicit coordinate dependence. Consequently, $\cF(\Phi,\Sigma;x)$ is no longer a purely superfield-built composite, and supersymmetry is generically obstructed by terms proportional to $\partial_x \cF$.

\subsection*{V.A. Breakdown of the Bogomol'nyi structure}

The breakdown of the Bogomol'nyi structure becomes particularly transparent in the bosonic static sector. Restricting, for simplicity, to the case \(\cF_\sigma=0\), and after eliminating the matter auxiliary field \(F\) and imposing the half-BPS spurion condition, we find
\begin{equation}
F(\phi,x)\equiv W_\phi(\phi)+\sigma(x)\,\partial_\phi \cF(\phi;x),
\label{eq:A-explicit-x}
\end{equation}
the static energy density takes the form
\begin{equation}
\cE
=
\frac12\,\phi'^2
+\frac12\,F(\phi,x)^2
+\eta\,\sigma'(x)\big(\cF(\phi;x)+\sigma(x)\big),
\label{eq:E-explicit-x}
\end{equation}
where, as before, $\eta=\pm1$ labels the preserved supercharge branch. Introduce the natural boundary functional
\begin{equation}
\cU(\phi,\sigma;x)\equiv W(\phi)+\sigma(x)\,\cF(\phi;x)+\frac12\,\sigma(x)^2.
\label{eq:U-explicit-x}
\end{equation}
Unlike the $x$-independent case, its derivative contains an additional term,
\begin{equation}
\frac{\dd\, \cU}{\dd x}
=
\phi'\,F(\phi,x)
+\sigma'(x)\big(\cF(\phi;x)+\sigma(x)\big)
+\sigma(x)\,\partial_x \cF(\phi;x).
\label{eq:Uprime-explicit-x}
\end{equation}
Therefore, completing the square yields
\begin{equation}
\cE
=
\frac12\Big(\phi'-\eta\,F(\phi,x)\Big)^2
+\eta\,\frac{\dd\, \cU}{\dd x}
-\eta\,\sigma(x)\,\partial_x \cF(\phi;x).
\label{eq:bogomolnyi-explicit-x}
\end{equation}
The last term is \emph{not} a total derivative in general. Hence an explicit coordinate dependence of $\cF$ spoils the simple Bogomol'nyi decomposition into a perfect square plus a boundary contribution, and correspondingly the sharp BPS bound is not protected in the same way. Only in special cases---for instance, if $\partial_x \cF(\phi;x)=0$ identically, or if $\sigma\,\partial_x \cF$ can be reorganized into a pure derivative by a model-dependent redefinition of $\cU$ --- does one recover a genuine BPS structure.

\subsection*{V.B. How to preserve half-supersymmetry:\\ promote $x$-dependence to a spurion}
From the superspace viewpoint the obstruction is conceptually clear: any ``hard'' coordinate dependence introduced by hand is not generated by a superfield background and is therefore not automatically compatible with a rigid (or partially preserved) supersymmetry algebra. The robust way to introduce spatial dependence while retaining control over supersymmetry is to generate it through \emph{spurion superfields}. Concretely, instead of $\cF(\Phi,\Sigma)\longrightarrow \cF(\Phi,\Sigma;x)$, one should consider
$
\cF(\Phi,\Sigma;x)\longrightarrow \cF(\Phi,\Sigma,\Xi),
\label{eq:K-to-Xi}
$
where $\Xi$ is an additional background superfield whose lowest component is fixed to the desired profile $\xi(x)$, while its fermionic and auxiliary components are turned on so that the background is invariant under a projected supercharge, i.e.\ by imposing the analogue of $\delta(\text{spurion fermion})=0$. This is precisely the logic already implemented by $\Sigma$: the ``compensating'' auxiliary data required to preserve a half-BPS subsector are not added ad hoc in components but arise systematically from the superspace completion of the background. In this spurionic realization the would-be $\partial_x \cF$ obstructions are replaced by supersymmetry-covariant background conditions, and one may recover a controlled half-BPS sector together with an exact Bogomol'nyi-type structure, now in an enlarged spurion space.

It is also interesting to note a potential parallel with recent multi-impurity constructions, where two independent profiles $\sigma_{1}(x)$ and $\sigma_{2}(x)$ are introduced to deform the first-order system in a two-field model \cite{Liao}. From the present spurion viewpoint, this structure may be viewed as suggestive of a multi-spurion extension, in which each impurity profile is promoted to a background superfield. While we do not pursue this connection here, it provides a natural avenue for relating multi-impurity parametrizations to a superspace embedding.

{{\bf VI.} {\bf \textit{Comments on derivative-dependent impurity couplings.}}
\label{sec:derivativeK}

Throughout this work, we have restricted our attention to impurity couplings of the form $\Sigma\,\cF(\Phi,\Sigma)$, namely, to the class in which $\cF$ depends only on the superfields $\Phi$ and $\Sigma$ themselves, and not on their supercovariant derivatives. This choice is minimal in the sense that it preserves the standard off-shell structure of the matter multiplet and leads to an auxiliary sector that can be eliminated algebraically without modifying the kinetic terms. It is natural, however, to ask what changes if the impurity functional is allowed to depend on superspace derivatives of $\Phi$. Schematically, one may consider
\begin{eqnarray}
S_{\rm imp}
=-\! \int\! \dd^2x\,\dd^2\theta\,\Sigma\,
\cF\!\left(\Phi,D^2\Phi,\Box\Phi,(\partial^M\Phi)^2,\ldots\right),
\label{eq:F-general}
\end{eqnarray}
where $\mathcal{F}$ is a real local functional built from $\Phi$ and its supercovariant and spacetime derivatives.

\subsection*{VI.A. Universality of the half-BPS spurion condition}

A key structural point is that the half-BPS condition on the spurion background remains \emph{universal}. Indeed, the requirement that a bosonic spurion profile preserve a supercharge is entirely controlled by the variations $\delta\lambda_\pm$, cf.\ \eqref{eq:delta-lp}--\eqref{eq:delta-lm}, and therefore depends only on the background fields $(\sigma,G)$ and their gradients. The details of how $\Sigma$ couples to the matter sector are irrelevant at this stage. Consequently, for static backgrounds one still finds
\begin{equation}
G(x)=\eta\,\sigma'(x),
\qquad
\varepsilon^- = \,\eta\,\varepsilon^+,
\qquad
\eta=\pm1,
\label{eq:spurion-universal}
\end{equation}
as the necessary and sufficient condition for the existence of a nontrivial projected supersymmetry.

\subsection*{VI.B. Backreaction on the matter auxiliary structure\\ and BPS equation}

In contrast, allowing $\cF$ to depend on derivatives of $\Phi$ generally modifies the way the matter auxiliary field $F$ enters the component Lagrangian. If the impurity functional depends explicitly on supercovariant or spacetime derivatives of the matter superfield,
\begin{equation}
\mathcal{F}=\mathcal{F}\!\left(\Phi,D_\pm\Phi,\partial_{\pm\pm}\Phi,\ldots\right),
\end{equation}
its $\theta$-expansion generically produces component terms in which derivatives act on $F$. This follows from the fact that
\begin{equation}
\partial_{\pm\pm}\Phi
\propto
\theta^+\theta^-\,\partial_{\pm\pm}F,
\end{equation}
so that any dependence of $\mathcal{F}$ on $\partial_{\pm\pm}\Phi$ may generate contributions proportional to $\sigma\,\partial_{\pm\pm}F$ in the $\theta^+\theta^-$ component of $\Sigma\,\mathcal{F}$.

However, the presence of derivative dependence does not by itself imply that $F$ becomes dynamical. In many cases, derivatives of $F$ appear only linearly and can be removed by integration by parts. Schematically, one encounters terms of the form
\begin{equation}
\int\dd^2x\;\sigma\,V^{\pm\pm}\,\partial_{\pm\pm}F
=
-\int\dd^2x\;\partial_{\pm\pm}(\sigma\,V^{\pm\pm})\,F,
\end{equation}
up to boundary terms, where $V^{\pm\pm}$ denotes a vector function. In such situations, no independent kinetic term for $F$ is generated, and the auxiliary field remains algebraic, although its equation of motion is deformed by coefficients involving the impurity profile and, in general, its gradients. What matters, therefore, is not merely the presence of derivatives in the impurity functional, but whether the chosen coupling preserves the algebraic character of the $F$-equation or instead generates genuine higher-derivative dynamics in the auxiliary sector.

The situation changes for genuinely higher-derivative impurity functionals, for instance when $\mathcal{F}$ is nonlinear in $\partial_{\pm\pm}\Phi$. In that case, the component action may contain terms quadratic or higher in derivatives of $F$, schematically of the form
$
\sigma(x)\,\big(\partial_{\pm\pm}F\big)^2+\cdots,
$
so that $F$ ceases to be purely auxiliary. Then its elimination becomes strongly model-dependent and may obstruct a simple closed-form BPS reduction and Bogomol'nyi completion. This is precisely why the class $\Sigma\,\cF(\Phi,\Sigma)$ considered in the main text is particularly distinguished: it preserves the standard off-shell auxiliary-field structure while still allowing for a nontrivial impurity background.

At the same time, there exist special subclasses of derivative-dependent couplings for which the auxiliary-field equation remains exactly solvable. As an example, let us consider a generalized impurity coupling of the form $\mathcal{F}=\mathcal{F}(\Phi,D^2\Phi,\Sigma)$. In this case, the bosonic contribution to the $\theta^2$ component of the impurity sector is
\begin{eqnarray}
\big[\Sigma\,\cF(\Phi,D^2\Phi,\Sigma)\big]_{\theta^2}\!\!
&=&
G\,\cF+\sigma\,\cF_\phi\,F \nonumber\\
&&+\cF_\sigma\,G+\sigma\,\cF_F\,\Box\phi,
\label{eq:SigKtheta}
\end{eqnarray}
where, at the component level, $\cF=\cF(\phi,F,\sigma)$. This expression already shows that, depending on how the dependence on $D^2\Phi$ is encoded in $\cF$, the resulting equation of motion for the auxiliary field $F$ may become nonlinear and strongly model-dependent, and in sufficiently general cases it may even be transcendental.

A particularly simple and still nontrivial subclass is obtained by choosing $\cF(\phi,F,\sigma)=F\,f(\phi,\sigma)$. In this case, the off-shell component Lagrangian becomes
\begin{eqnarray}
\cL_{\rm off}
&=&\;
\frac12\,\phi\,\Box\phi
+\frac12\,F^2\Big(1-2\sigma f_\phi \Big)
\nonumber\\
&&-F\Big(W_{\phi}+G\,f+G\,\sigma\,f_{\sigma}\Big)
\nonumber\\
&&-\sigma\Big(G+f\,\Box\phi\Big)
+\ldots,
\label{eq2:Loff-full}
\end{eqnarray}
where the ellipsis denotes the fermionic sector. The corresponding equation of motion for the auxiliary field is then readily obtained as
\begin{eqnarray}
F=\frac{W_\phi(\phi)+G\,(f(\phi,\sigma)+\,\sigma\,f_\sigma(\phi,\sigma))}{\kappa(\phi,\sigma)},
\label{eq2:F-eom}
\end{eqnarray}
where we have introduced the shorthand
\begin{eqnarray}
\kappa(\phi,\sigma)\equiv 1-2\sigma(x) f_\phi(\phi,\sigma).
\label{eq:kappa-def}
\end{eqnarray}
Therefore, although derivative-dependent impurity couplings generically complicate the auxiliary-field structure, this example shows that suitable subclasses may still preserve exact solvability.

In this special subclass, it is still possible to derive a first-order equation of Bogomol'nyi type and to rewrite the static energy in a correspondingly deformed Bogomol'nyi form. To this end, let us consider the bosonic static sector and impose the half-BPS spurion condition Eq.\eqref{eq:spurion-universal}.
Moreover, because the supersymmetry variation of the matter multiplet is unchanged, Eq.~\eqref{eq:F-phi} continues to hold in the bosonic BPS sector.
Substituting \eqref{eq:spurion-universal} into \eqref{eq2:F-eom}, and using \eqref{eq:F-phi}, one finds the generalized Bogomol'nyi equation
\begin{eqnarray}
\kappa(\phi,\sigma)\,\phi'
=
\eta\,W_\phi+\sigma'\,\big(f+\sigma f_\sigma\big).
\label{eq:BPS-gen-der}
\end{eqnarray}

Let us now turn to the static energy density. Starting from \eqref{eq2:Loff-full}, restricting ourselves to the bosonic static sector, and integrating by parts the term proportional to $f\,\Box\phi$, one obtains
\begin{eqnarray}
\cE
&=&
\frac12\,\kappa(\phi,\sigma)\,\phi'^2
-\frac12\,\kappa(\phi,\sigma)\,F^2
+\sigma G
\nonumber\\
&+&
\!\! F\Big[W_\phi+G\,f+G\,\sigma f_\sigma\Big]
-\sigma'\big(f+\sigma f_\sigma\big)\phi'
.
\label{eq:E-static-gen1}
\end{eqnarray}
Upon imposing \eqref{eq:spurion-universal}, eliminating the auxiliary field via \eqref{eq2:F-eom}, and completing the square, the static energy can be cast in the form
\begin{eqnarray}
\cE
&=&
\frac12\,\kappa(\phi,\sigma)
\left[
\phi'
-
\frac{\eta W_\phi+\sigma'(f+\sigma f_\sigma)}
{\kappa(\phi,\sigma)}
\right]^2
\nonumber\\
&&
+\eta\,\frac{d}{dx}
\left(
W(\phi)+\frac12\,\sigma^2
\right).
\label{eq:E-static-BPS-gen}
\end{eqnarray}
Therefore, the static energy takes a deformed Bogomol'nyi form, and the corresponding bound is saturated precisely when \eqref{eq:BPS-gen-der} holds. In particular, provided that
$\kappa(\phi,\sigma)>0$, 
the square term is non-negative and one obtains the bound
\begin{eqnarray}
E\geq |\Delta U|,
\qquad
U(\phi,\sigma)=W(\phi)+\frac12\,\sigma^2.
\label{eq:bound-gen}
\end{eqnarray}
Thus, although the derivative-dependent coupling deforms both the first-order equation and the quadratic structure of the energy density, this particular subclass still preserves an exact Bogomol'nyi-type construction.

{{\bf VII.} {\bf \textit{Conclusions.}}}
\label{sec:conclusions}
In this work we provided a rigid $\mathcal{N}=(1,1)$ superspace completion of impurity deformations in $D=1+1$ dimensions by promoting the prescribed impurity profile $\sigma(x)$ to a real spurion superfield $\Sigma$. This spurionic embedding makes the supersymmetry structure fully transparent and allows one to identify, without any \emph{ad hoc} component-level adjustments, the precise circumstances under which a spatially varying impurity preserves a nontrivial subset of supercharges. In our view, this framework appears to be very efficient for generalization.

We established that preserving a nontrivial supercharge in a static bosonic spurion background imposes a necessary and sufficient relation between the spurion components, which simultaneously induces the corresponding half-BPS projector on the supersymmetry parameters and selects the preserved branch. In the resulting half-BPS sector, the projected vanishing of the matter-fermion variations, combined with the algebraic elimination of the auxiliary field, yields a universal first-order equation for static bosonic configurations. We further showed that the static energy admits an exact Bogomol'nyi completion into a perfect square plus a total derivative, leading to a sharp bound governed by a boundary functional and saturated precisely by solutions of the BPS equation.

We also clarified how introducing explicit coordinate dependence in the impurity functional generically spoils the simple Bogomol'nyi structure by producing non-derivative obstruction terms, and we argued that the robust way to retain supersymmetric control over inhomogeneities is to generate them through additional spurion backgrounds. Moreover, allowing derivative-dependent impurity couplings typically backreacts on the auxiliary-field structure and may obstruct an algebraic elimination of auxiliaries; a systematic classification of derivative-dependent functionals that still support an algebraic auxiliary sector, a closed first-order BPS reduction, and an exact Bogomol'nyi bound is an interesting direction for future work. Other possibilities concern working with an enlarged multi-field system similar to the two-field model considered in \cite{Liao} and also, with higher-dimensional spacetimes, where one can consider impurity couplings to localized structures such as vortices \cite{V1,V2,V3} and magnetic monopoles \cite{Wei,Gia,Bazeia:2025edo}.

\vspace{.5cm}
{\bf Acknowledgments:}
The authors acknowledge partial financial support from the Conselho Nacional de Desenvolvimento Cient\'ifico e Tecnol\'ogico (CNPq) under Grants No.~303469/2019-6 (DB), No.~402830/2023-7 (DB), No.~404310/2023-0 (ACL), and No.~301256/2025-0 (ACL). They also thank Gabriel~S.~Santiago for useful comments.

{\bf Data Availability Statement:} This manuscript has no associated data. [Author’s comment: Data sharing not applicable to this article as no datasets were generated or analysed during the current study.]

{\bf Code Availability Statement:} This manuscript has no associated
code/software. [Author’s comment: Code/Software sharing not applicable to this article as no code/software was generated or analysed during
the current study.]

\appendix*
\section{Conventions and notations}
\label{sec:conventions}

The superspace structure in two and three spacetime dimensions is closely parallel; accordingly, we adopt notation and conventions that follow Refs.~\cite{GGRS,Shizuya:2003vm} as closely as possible. For the reader's convenience, we collect here the definitions and conventions used throughout the paper, organized as follows. We first fix our spacetime conventions and light-cone operators. We then summarize our gamma-matrix representation and the associated Majorana--Weyl spinor conventions. Finally, we state our $\mathcal{N}=(1,1)$ superspace conventions, including the normalization of superspace measures and the definitions of supercharges and supercovariant derivatives.

\subsection{Spacetime, spinors and gamma matrices}

We work in flat $D=1+1$ Minkowski spacetime with signature $(-,+)$ and Cartesian coordinates $(t,x)$, following as closely as possible the conventions of Ref.~\cite{GGRS}. The proper orthochronous Lorentz group is $SO^+(1,1)$, whose fundamental representation acts on real Majorana spinors $\psi^\alpha=(\psi^+,\psi^-)$. Spinor indices are denoted by Greek letters, $\alpha,\beta=\pm$, whereas spacetime indices are denoted by capital Latin letters, $M,N=0,1$. A spacetime vector $x_M$ is equivalently represented by a symmetric rank-two spinor,
$$
x^{\alpha\beta}=(\gamma^M)^{\alpha\beta}x_M,
\qquad
x^{\alpha\beta}=x^{\beta\alpha},
$$
whose independent components may be denoted as $(x^{++},x^{--})$.

Spinor indices are lowered and raised with the antisymmetric charge-conjugation matrix
\begin{equation}
C_{\alpha\beta}=-C_{\beta\alpha}=-C^{\alpha\beta}=
\begin{pmatrix}
0 & -i\\
i & 0
\end{pmatrix},
\label{eq:Cab}
\end{equation}
so that
\begin{eqnarray}
\psi_\alpha &=& \psi^\beta C_{\beta\alpha},
\qquad
\psi^\alpha = C^{\alpha\beta}\psi_\beta, \nonumber\\[2pt]
\psi^2 &\equiv& \frac12\,\psi^\alpha\psi_\alpha
= i\,\psi^+\psi^-.
\end{eqnarray}

We employ the following representation of the Clifford algebra
\begin{equation}
\{\gamma^M,\gamma^N\}=2\eta^{MN},
\qquad
\eta^{MN}=\mathrm{diag}(-,+),
\label{eq:clifford}
\end{equation}
given by
\begin{equation}
{(\gamma^0)^{\alpha}}_{\beta}=\sigma^2=
\begin{pmatrix}
0 & -i\\
i & 0
\end{pmatrix},
\quad
{(\gamma^1)^{\alpha}}_{\beta}=-i\sigma^1=
\begin{pmatrix}
0 & -i\\
-i & 0
\end{pmatrix}.
\label{eq:gamma-matrices}
\end{equation}

In terms of these notations, the kinetic contraction reads
\begin{equation}
\partial_M\phi\,\partial^M\phi
=-(\partial_t\phi)^2+(\partial_x\phi)^2=\frac{1}{2}\partial^{\alpha\beta}\phi\,\partial_{\alpha\beta}\phi
\label{eq:kin-ppmm}
\end{equation}
For static profiles $f=f(x)$ one has
\begin{equation}
\partial_{++} f = + f'(x),
\qquad
\partial_{--} f = - f'(x).
\label{eq:static-ppmm}
\end{equation}

\subsection{$\cN=(1,1)$ superspace}

Rigid $\cN=(1,1)$ superspace is parametrized by $(x^{\pm\pm},\theta^\pm)$ with real Grassmann coordinates $\theta^\pm$. We define supercharges and supercovariant derivatives by
\begin{equation}
Q_\pm=\frac{\partial}{\partial\theta^\pm}-i\,\theta^\pm\,\partial_{\pm\pm},
\,
D_\pm=\frac{\partial}{\partial\theta^\pm}+i\,\theta^\pm\,\partial_{\pm\pm},
\label{eq:Q-D}
\end{equation}
so that $
\{Q_\pm,Q_\pm\}=-2i\,\partial_{\pm\pm}
$,
$\{D_\pm,D_\pm\}=+2i\,\partial_{\pm\pm},$ 
and 
$\{Q_\pm,D_\pm\}=0,$
\label{eq:algebra}
with all other (anti)commutators vanishing. Moreover, supersymmetry variations are such that $\delta=\varepsilon^+Q_+ + \varepsilon^-Q_-$ with constant real Grassmann parameters $\varepsilon^\pm$, and we are considering Berezin integration normalized as
\begin{equation}
\int \dd^2\theta\;\theta^2 =- 1.
\label{eq:berezin}
\end{equation}



\begin{thebibliography}{99}
\bibitem{Malomed}Y. S. Kivshar and B. A. Malomed, Rev. Mod. Phys. 61, 763 (1989) doi:10.1103/RevModPhys.61.763.
\bibitem{impu}Y. S. Kivshar, Z. Fei and L. Vazquez, Phys. Rev. Lett. 67, 1177 (1991) doi:10.1103/PhysRevLett.67.1177.

\bibitem{Adam} C. Adam, K. Oles, T. Romanczukiewicz and A. Wereszczynski, Phys. Rev. Lett. 122, 241601 (2019) doi:10.1103/PhysRevLett.122.241601 [arxiv:1903.12100 [hep-th]].
\bibitem{B}E. B. Bogomol’nyi, Sov. J. Nucl. Phys. 24, 449 (1976).
\bibitem{PS}M. K. Prasad and C. M. Sommerfield, Phys. Rev. Lett.
35, 760 (1975).
\bibitem{Bazeia:2023rug}
D.~Bazeia, M.~A.~Liao and M.~A.~Marques,
Phys. Lett. B {846}, 138262 (2023)
doi:10.1016/j.physletb.2023.138262
[arXiv:2310.09862 [hep-th]].

\bibitem{Bazeia:2023dyr}
D.~Bazeia, M.~A.~Liao and M.~A.~Marques,
Eur. Phys. J. C {84}, 180 (2024)
doi:10.1140/epjc/s10052-024-12510-5
[arXiv:2311.14661 [gr-qc]].

\bibitem{Adam:2019yst}
C.~Adam, J.~M.~Queiruga and A.~Wereszczynski,
J. High Energ. Phys. 2019, 164 (2019)
doi:10.1007/JHEP07(2019)164
[arXiv:1901.04501 [hep-th]].

\bibitem{GGRS} S.~J.~Gates, M.~T.~Grisaru, M.~Ro\v{c}ek, and W.~Siegel, \textit{Superspace, or One Thousand and One Lessons in Supersymmetry} (Benjamin/Cummings, 1983). 

\bibitem{Shizuya:2003vm}
K.~Shizuya,
Phys. Rev. D {69}, 065021 (2004)
doi:10.1103/PhysRevD.69.065021
[arXiv:hep-th/0310198 [hep-th]].

\bibitem{Liao}D. Bazeia, M. A. Liao, M. A. Marques, Chaos, Solitons \& Fractals 192, 115950 (2025) doi:10.1016/j.chaos.2024.115950 [arxiv:2409.03603 [hep-th]].

\bibitem{V1}D. Tong and K. Wong, J. High Energ. Phys. 2014, 90 (2014)
doi:10.1007/JHEP01
 [arXiv:1309.2644 [hep=th]].

\bibitem{V2}X. Han and Y. Yang, J. High Energ. Phys. 2016, 46 (2016) doi:
https://doi.org/10.1007/JHEP02
[arXiv:1510.07077 [hep-th]].

\bibitem{V3}D. Bazeia, J. G. F. Campos and A. Mohammadi, J. High Energ. Phys. 2024, 108 (2024) doi:10.1007/JHEP12(2024)108 [arXiv:2404.11694 [hep-th]]

\bibitem{Wei}E. J. Weinberg, {\it Classical Solutions in
Quantum Field Theory} (Cambridge University Press, 2014).

\bibitem{Gia}M. Bachmaier, G. Dvali, J. Seitz, J. S. Valbuena-Bermúdez, Phys. Rev. D 111, 075014 (2025) doi:10.1103/PhysRevD.111.075014 [arxiv:2502.01756 [hep=th]].

\bibitem{Bazeia:2025edo}
D.~Bazeia, M.~A.~Liao and M.~A.~Marques, {\it Magnetic monopoles in Yang-Mills-Higgs theory with impurities}
[arXiv:2505.11215 [hep-th]].

\end{thebibliography}
\end{document}